\def\RE {I\kern-6pt R    }
\def\Z  {Z\kern-13pt Z   }
\def\be {\begin{equation}}
\def\ee {\end{equation}  }
\def\beq{\begin{eqnarray}}
\def\eeq{\end{eqnarray}  }
\def\lb{\left(}
\def\rb{\right)}
\def\bi {\begin{itemize} }

\def\ei {\end{itemize}   }

\documentstyle[twocolumn,prd,aps]{revtex}

\input epsf

\begin{document}
\draft

\twocolumn[\hsize\textwidth\columnwidth\hsize\csname
@twocolumnfalse\endcsname

\title{Pair Production in the Collapse of a Hopf Texture}

\author{Eric W. Hirschmann and Steven L. Liebling}
\address{Center for Relativity,
         The University of Texas at Austin,
         Austin, TX 78712-1081}

\maketitle

\begin{abstract}

We consider the collapse of a global ``Hopf" texture and examine
the conjecture, disputed in the literature,
that monopole-antimonopole pairs can be formed in
the process.  We show that such monopole-antimonopole pairs can indeed
be nucleated in the course of texture collapse given appropriate initial
conditions.  The subsequent dynamics
include the recombination and annihilation of the pair in a burst of
outgoing scalar radiation.

\end{abstract}

\vskip2pc]

\section{Introduction}
\label{sec:tex_introduction}

An outstanding cosmological problem is explaining the homogeneity 
and isotropy of the universe on large scales while allowing for 
sufficient inhomogeneity and anisotropy on small scales to account 
for the existence
of galaxies, stars, and, of course, us.  Several mechanisms for this
have been suggested.  One of these is the inflationary 
scenario.  An alternative which has also received considerable 
attention is the idea that the production of topological defects via phase
transitions in the early universe provided the density fluctuations
which eventually led to galaxy formation and the like.  

In general, defects result because of a spontaneously broken symmetry. 
If one begins with a set of fields, $\Phi^a$, which possess a 
global symmetry $G$
and interact via a potential $V(\Phi)$ which breaks $G$ down to a 
subgroup $H$, the vacuum manifold is described by the quotient space
$G/H$.  Defects will arise in theories for which the vacuum manifold 
allows nontrivial homotopy groups $\pi_n(G/H)$
characterized by an integer $n$. 
The homotopy groups serve to
differentiate mappings from the $n$-dimensional sphere $S^n$
into the manifold $\cal M$~\cite{vilenkin}.
For different  values of $n$,
one gets different types of defects.  Domain walls arise for $n=0$, 
global strings for $n=1$, monopoles for $n=2$, and textures for $n=3$.  

At very early times, the universe is at a very high temperature
so that, in general, the effective potential has a unique, symmetric
vacuum state which does not allow defects.
However, as the universe expands,
it cools and passes through a symmetry breaking phase transition allowing
for the possible formation of these defects. The vacuum manifold
is no longer a single unique state, but instead a collection of 
degenerate vacuum 
states which comprise a 
nontrivial manifold.
The presence of these defects can then serve as gravitational seeds
for structure formation, and hence the importance of understanding
their dynamics.

Here, we will be interested in a particular texture model and its
dynamics. 
In accordance with Derrick's theorem, scalar field configurations
corresponding to a texture are unstable and will tend to shrink or collapse~\cite{vilenkin}.
An interesting conjecture has recently been put forth on the collapse
of various texture models.  Sornborger {\it et al}~\cite{sornborger3} suggest that the 
nature of the various collapses can be categorized by whether the 
particular texture model allows more than one type of defect.  In 
particular, models for which only $\pi_3$ is
nontrivial (such as $SO(4)/SO(3)$) 
collapse in a similar (and, from their results, nearly
indistinguishable) manner.  On the other hand, models for which homotopy 
groups $\pi_3$ and $\pi_2$ are both nontrivial (such as $SO(3)/SO(2)$)  
should collapse in a qualitatively different manner, with the possible 
production of defects characterized by the
other (non-$\pi_3$) nontrivial homotopy
group.  These authors provide numerical evidence for their conjecture,
having evolved several texture models from similar initial conditions
and examined the collapse process/products.  

The results in~\cite{sornborger3} would appear to be 
in agreement with interesting experimental evidence in 
nematic liquid crystals. 
Chuang {\it et al}~\cite{chuang} 
induce phase transitions in a nematic liquid crystal
described by  a $SO(3)/SO(2)$ broken symmetry  and produce textures.
These textures, they claim, decay
via the production of a monopole-antimonopole
pair.  However, later numerical work by 
Rhie and Bennett~\cite{rhie} 
with this model suggests that such pair production
did not occur.  Later still, Luo~\cite{luo} suggests that the
reason for the absence of pair production in the numerical simulations
of Rhie and Bennett
is that the initial configuration of their simulations
does not correspond to that seen in
the laboratory experiments of Chuang. Instead, he presents different initial
data which he claims should indeed nucleate monopoles.

In the following, we analyze the dynamics of collapsing Hopf textures,
those textures  occurring
from the breaking of $SO(3)$ to $SO(2)$.
The results verify aspects of
these earlier results and clarify the general picture.  In the
first section, we present the model, the evolution equations, and discuss
the numerical approach.  In subsequent sections, we consider the evolutions
of a variety
of initial data.  These include a single monopole and a 
monopole-antimonopole pair which are evolved to demonstrate that the code can 
evolve these objects and to allow   the identification of
their presence or absence
in later evolutions of texture collapse.

\section{The Equations and Numerical Approach}
\label{sec:tex_eqns}
The model we consider is described by the Lagrangian 
\be
{\cal L} = -\frac{1}{2} \Phi^a_{,\mu} \Phi^{a,\mu} - V(\Phi)
\ee
where the potential $V(\Phi)$ is
\be
V(\Phi) = \gamma \left( \Phi^a \Phi^a - v^2 \right)^2
\ee
and $\Phi^a$ is a set of three scalar fields.  The scalar
fields thus transform under the global symmetry group $SO(3)$ with
the potential breaking this down to $SO(2)$.  The resulting vacuum
manifold is $SO(3)/SO(2) = S^2$ for which the second and third 
homotopy groups are nontrivial.  Thus this model allows both textures  
and global monopoles.  The equations of motion are
\be
\label{eom}
\partial_\mu \partial^\mu \Phi^a = 4\gamma \Phi^a 
			\left(\Phi^b\Phi^b - v^2 \right).
\ee
A trivial rescaling invariance $\Phi^a \rightarrow v\,\Phi^a$
allows us to set $v=1$, leaving the model with one free,
dimensionless parameter $\gamma$.

Following~\cite{sornborger3} we split the total energy density into three
parts:  the kinetic, gradient and potential energy densities defined
as
\beq  
\rho_k & = & {1\over2} \left( \dot{\Phi}^a \dot{\Phi}^a \right), \\ 
\rho_g & = & {1\over2} \left( \nabla\Phi^a \nabla\Phi^a \right), \\
\rho_v & = & \gamma \left( \Phi^a \Phi^a - 1 \right), 
\eeq 
where an overdot denotes derivative with respect to time
and $\nabla$ represents the spatial gradient. The total energy density is
\be
\rho_T = \rho_k + \rho_g + \rho_v.
\ee

This model also has a conserved topological charge.
The monopole current associated with the
charge density is
\be
k_\mu = \frac{1}{8\pi} \epsilon_{\mu \nu \rho \sigma} \, \epsilon_{abc} \,
           \phi^a_{,\nu} \, \phi^b_{,\rho} \, \phi^c_{,\sigma},
\ee
where $\epsilon_{\mu \nu \rho \sigma}$ and $\epsilon_{abc}$
are purely antisymmetric tensors 
of rank $4$ and $3$ respectively.
The conserved monopole charge is
\be
Q = \int k_0~d^3x = \frac{1}{8\pi} \int d^3 x \left(
                       \epsilon_{i j k} \, \epsilon_{abc} \,
           \phi^a_{,i} \, \phi^b_{,j} \, \phi^c_{,k}
              \right),
\ee
the integral of the charge density $k_0$~\cite{raj}.

Given appropriate initial conditions, these equations are now 
straightforward to integrate in Cartesian coordinates.  Some particular
initial data that has been suggested for texture collapse is
\beq 
\label{carrolldata}
\Phi(\vec x) = \left( \begin{array}{c}
    \cos^2 \lb \chi \rb + \bigl( 2 z^2 / r^2 - 1\bigr) \sin^2 \lb \chi \rb \\
      2\bigl( x z \sin^2 \lb \chi \rb / r^2 
             +  y \cos \lb \chi \rb \sin \lb \chi \rb / r \bigr) \\
      2\bigl( y z \sin^2 \lb \chi \rb / r^2 
  	     -  x \cos \lb \chi \rb \sin \lb \chi \rb / r \bigr) \\
                                             \end{array} \right),
\eeq  
where $r = \sqrt{x^2+y^2+z^2}$ is the usual spherical radial coordinate 
and where the boundary conditions on the radial function $\chi(r)$ are
$\chi(0) = 0$ and $\chi(\infty) = \pi$. 
We discuss the evolution of this initial data as well as the evolution
of another form of initial data for a texture in Sections~\ref{sec:sph}
and~\ref{sec:tor}.

It is worth noting that for the initial data in Eq.(\ref{carrolldata}),
the initial total energy 
density is axisymmetric.  For the particular case
in which $\chi(r) = 2\tan^{-1} \left(r\right)$, the initial energy
is actually spherically symmetric. In addition, one may expect this axisymmetry
to be maintained during the evolution.  We attempt to take advantage of 
this observation and cast the equations
and this initial data into a manifestly axisymmetric form.  In
cylindrical coordinates ($\rho,z,\varphi$), we define new quantities
\beq  
\Phi^{+} & = & \cos\varphi \: \Phi^2 + \sin\varphi \: \Phi^3 \\ 
\Phi^{-} & = & \sin\varphi \: \Phi^2 - \cos\varphi \: \Phi^3 
\eeq
which, for the above initial data, can be seen to be manifestly 
axisymmetric.  It is now a simple matter to get the equations of motion
for $\Phi^{\pm}$.  Using the equations of motion for $\Phi^{a}$ from 
Eq.(\ref{eom})  
we find the equations for $\Phi^{\pm}$ to be  
\beq  
-\Phi^{+}_{,tt} & + & {1\over\rho}\left(\rho\Phi^{+}_{,\rho}\right)_{,\rho}
     - {1\over \rho^2} \Phi^{+}
     + \Phi^{+}_{,zz} + {2\over\rho^2} \Phi^{-}_{,\varphi}
     + {1\over \rho^2} \Phi^{+}_{,\varphi\varphi}        \nonumber \\
 & = & 4\gamma \Phi^{+}(\Phi^{b}\Phi^{b} - v^2)  \\ 
-\Phi^{-}_{,tt} & + & {1\over\rho}\left(\rho\Phi^{-}_{,\rho}\right)_{,\rho}
     - {1\over \rho^2} \Phi^{-}
     + \Phi^{-}_{,zz} - {2\over\rho^2} \Phi^{+}_{,\varphi}
     + {1\over \rho^2} \Phi^{-}_{,\varphi\varphi}        \nonumber \\
  & = &  4\gamma \Phi^{-}(\Phi^{b}\Phi^{b} - v^2)   
\eeq
with
\beq 
\Phi^{b}\Phi^{b} 
    & = & \Phi^{1}\Phi^{1} + \Phi^{+}\Phi^{+} + \Phi^{-}\Phi^{-} .
\eeq  
Since the evolution will remain axisymmetric, 
we can discard the terms which have
derivatives with respect to $\varphi$.

This simplification of the problem from 3D to 2D results in enormous
gains from a computational perspective and allows us significant improvement
in our potential resolution of dynamic features in the evolution.  However,
in our code development and tests, we found it very useful to also have a 3D 
code with which to compare our 2D results and results of earlier investigators.

 For those reasons, we have implemented both a 3D code and a 2D  
axisymmetric code using
a second order Crank-Nicholson finite difference scheme implemented
with RNPL~\cite{marsa}.
Out-going radiation conditions based on approximate spherical
propagation 
as $r\rightarrow \infty$ are imposed
\be
r\Phi^a = f(t-r),
\ee
for some function $f$.
This condition
is very effective in limiting reflection from the boundaries. We 
have verified that
these codes are fully convergent and that they are stable for many
crossing times.

\section{Single Monopole}
\label{sec:tex_monopole}
As stated in the introduction, we want to evolve a collapsing Hopf texture
and examine the possible nucleation of monopole-antimonopole pairs.
However, in order to understand the behavior of any nucleated
monopoles modeled by these codes,
we first examine the evolution of initial data explicitly containing
a monopole.
Using the standard
hedgehog ansatz, we can represent a monopole of unit charge at the origin
by
\be
\Phi\left(\vec x\right) = \frac{f(r)}{r}
                          \left( \begin{array}{c}
                                 x \\
                                 y \\
                                 z
                                 \end{array} \right),
\label{eq:hedge}
\ee
where
\be
f(0)=0$ ~~~~ $f(\infty)= 1.
\label{eq:bc}
\ee
To find the static monopole $\tilde f(r)$,
we substitute the ansatz~(\ref{eq:hedge}) into
Eq.(\ref{eom}) and set all time derivatives to zero
\be
\tilde f'' + \frac{2\tilde f}{r} - \frac{2\tilde f}{r^2} = 4 \gamma \tilde f \left( \tilde f^2 - 1\right).
\ee
We then solve to find $\tilde f(r)$.
However, instead of using the exactly static solution, $\tilde f(r)$, which
is exact only in the continuum limit, we note that
the static monopole is rather well approximated by
\be
\tilde f(r) \approx \tanh \left( \sqrt{\gamma} r \right).
\ee
Using this as our initial data shows both the energy
and monopole charge to be well conserved.

Because the monopole is a topological object, we expect
that any initial data which has the same
boundary conditions will also have unit topological charge.
Hence, we can choose the function $f(r)$ different than
$\tilde f(r)$ but still obeying Eq.(\ref{eq:bc}) to set
our initial data. This data will not be static but will still
have a conserved monopole charge. As an example,
we examine the evolution of initial data of the form
\be
f(r) = \frac{1}{2} \left[ 1 + \tanh \left( \frac{r - R}{\beta} \right) \right]
\ee
with $R$ and $\beta$ arbitrary constants.

These simulations have been done
with time symmetric initial data.
However, we have also implemented a boost of the monopole
by giving it an initial velocity in some direction along an axis.
The monopole is seen traversing the numerical domain. That we can
evolve a monopole for many crossing times and also boost it while
retaining its particle nature, provides further evidence that
the global monopole is stable to radial perturbations as discussed in 
\cite{bennett,goldhaber,rhie2}.

\section{Monopole-Antimonopole Pair}  

Next, we would like to consider evolving a
monopole-antimonopole pair with our codes.
An anti-monopole is just a reflection of the monopole ansatz~(\ref{eq:hedge}).  Thus, it takes 
the form
\be
\Phi\left(\vec x\right) = \frac{f(r)}{r}
                          \left( \begin{array}{c}
                                  x \\
                                  y \\
                                 -z
                                 \end{array} \right).
\ee
Strictly speaking, there is not a superposition principle
for monopoles and anti-monopoles. However, it is reasonable to assume that if the
separation between the monopole and anti-monopole making up a pair
is considerably larger than the core width of the respective
particles, a pair configuration can be well approximated by 
matching a monopole solution to an anti-monopole solution.
To achieve this, we translate the monopole up and the antimonopole down some
distance $L$ to get the solution for a monopole-antimonopole pair
located on the $z$ axis
\be
\Phi\left(\vec x\right)  =
                           \frac{f}{\sqrt{x^2 + y^2 + \lb |z|-L \rb^2}}
                              \left( \begin{array}{c}
                                     x \\
                                     y \\
                                     |z| - L
                                     \end{array} \right),
\ee
where, in analogy with the previous section, we consider two forms for
$f$
\be
f = \tanh \left( \sqrt{\gamma \left( x^2 + y^2 + \left( |z| - L \right)^2 \right) } \right)
\ee
and
\be
f = \frac{1}{2} \left[ 1 + \tanh \left( \frac{\sqrt{x^2 + y^2 + \lb |z|-L \rb^2} - R}{\beta} \right) \right].
\label{eq:1+tanh}
\ee
At $z=0$, the fields $\Phi^a$, though not differentiable, are continuous.
This non-smoothness
gets smoothed quickly by the numerical evolution and does not appear
to affect the evolution of the pair significantly.

The total charge of this system during the evolution should be
identically zero. Indeed, we confirm this numerically as the total
charge remains essentially zero throughout the
evolution. Though a necessary check for our code, this fact makes
it a bit more difficult to follow the evolving charges individually.
One quantity which we do track is the topological charge density.  This, of
course, should be peaked at the locations of the monopoles.  However, we
also found it useful to define and calculate a quantity which we call
the ``half-space charge.''  It is simply the topological charge density integrated
over positive $z$ only
\be
Q_{1/2} =\int_x \int_y \int_{z=0}^{z_{\rm max}} ~k^0~dx~dy~dz.
        =\int_\rho \int_{z=0}^{z_{\rm max}} 2\pi\rho~k^0~d\rho~dz.
\ee
We emphasize that this quantity has no topological meaning and is not
necessarily conserved. However, as with our earlier argument for superposing
the monopole-antimonopole pair, if we imagine the constituents of
the pair to be well separated and individually well localized, the
half-space charge should approximate the charge of the monopole (or anti-monopole).

\begin{figure}
\epsfxsize=7.5cm
\centerline{\epsffile{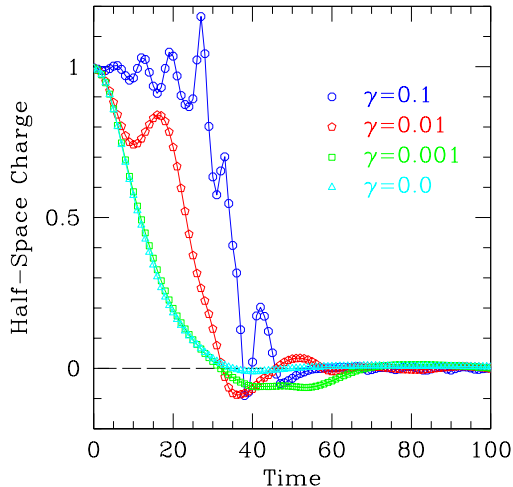}}
\caption[Half-space charge for a monopole-antimonopole pair.]
{The half-space charge
for the evolution of a monopole-antimonopole pair. The
monopoles are described by Eq.(\ref{eq:1+tanh}) with initial 
separation $2L=30$, $\beta = 3.5$, $R = 10$ on a 257x513 grid
with outer boundaries at $z_{max}=-z_{min}=\rho_{max}=40$.   
For small $\gamma$, the half-space charge quickly goes to zero,
while for larger
$\gamma$ a well defined annihilation occurs releasing a large
amount of bosonic radiation. The charge for the other half-space
is simply the reflection of the graph about zero (the dashed line).
}
\label{fig:pair}
\end{figure}

We then evolve the pair configuration for $L=15$ and show the resulting
half-space charge in Fig.~\ref{fig:pair}. 
Note that for early times, the average value of this quantity for large enough $\gamma$
is roughly unity as we would expect for an isolated monopole.  For smaller 
$\gamma$, the half-space charge becomes zero much more quickly. 
This behavior would appear consistent with the fact that decreasing $\gamma$ has the
effect of increasing the size of the respective monopoles. The increased size of the core
makes it more difficult for us to associate the half-space charge
with the true topological charge and results in more rapid pair annihilation.

We have investigated the oscillations appearing in the half-space charge
seen in Fig.~\ref{fig:pair}.
The oscillations converge with increasing resolution. Further,
by moving the boundary farther from the monopole pair while keeping the
same effective resolution, the oscillations remain the same. 
Both these results indicate that the oscillations
are a real phenomenon and not a numerical artifact. Hence, the oscillations
appear to be part of the dynamics.  That this quantity oscillates is
not inconsistent with charge conservation as this half-space charge
is not a topological invariant. For it to be so, we would necessarily have
to extend the integral out to infinity which we clearly cannot do
while distinguishing the monopole from the pair. As further support
that these oscillations are representative of the dynamics,
oscillations of the  total potential energy  are observed
to be correlated to the oscillations in the
half-space charge.

\begin{figure}
\epsfxsize=7.5cm
\centerline{\epsffile{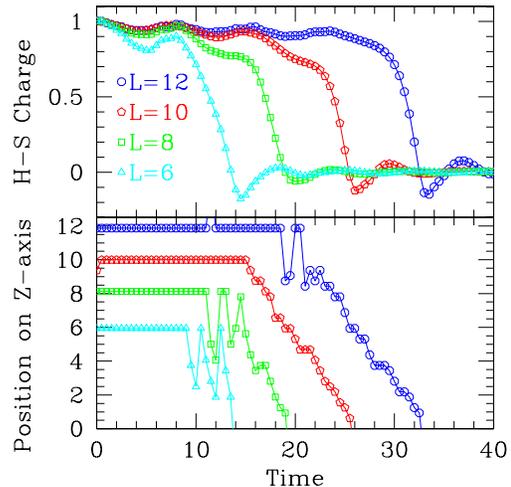}}
\caption{Evolutions of various separations $L$ for a monopole-antimonopole pair.
         The top graph shows the half-space charge. The lifetime
         of the pair is expectedly dependent on the separation.
         The bottom graph shows the position on the z-axis of the maximum
         charge density up to the time of annihilation.
         Taking this value to represent the location
         of the monopole, the graph shows the monopole accelerated
         towards annihilation after some time dependent on the initial
         separation.
         These were run on a 65x129 grid with $\gamma=0.1$. 
         Changes to $R$ and $\beta$ change
         these graphs very little (here, $R=1.5$ and $\beta=1.0$).
}
\label{fig:pair_L}
\end{figure}

In Fig.~\ref{fig:pair_L}, the results of varying the parameter $L$
in the pair configuration are shown. The graph of the half-space
charge shows that the lifetime of the pair is clearly proportional
to the initial separation. The bottom graph displays the position
on the $z$ axis
of the maximum charge density as a function of time. Because of 
our somewhat ad hoc matching procedure, the monopoles are not
influenced by each other until a sufficient amount of time has 
passed for a signal to have propagated between them.  
At that point, the monopoles experience an attractive force, 
approach each other and annihilate.

\section{``Spherically Symmetric" Texture} 
\label{sec:sph}
With a good idea of what the signature of a monopole-antimonopole
pair might be,
we now evolve the initial data given in Eq.(\ref{carrolldata}) with
\begin{equation}
\label{ourchi}
\chi(r) =
   \frac{\pi}{2} \left[ 1 + \tanh\left( \frac{r-R}{\beta} \right) \right].
\end{equation}
Though the precise shape of this function does not appear to 
significantly affect
the results, the function $\chi(r)$ used here differs from that
of~\cite{sornborger3} because the function shown there is not
smooth.

As we stated in Section~\ref{sec:tex_eqns}, the total energy density 
is spherically symmetric only for $\chi(r) = 2 \tan^{-1} (r)$.  We are therefore
being a bit careless to refer to this as spherically symmetric initial
data.  In addition, as shown in~\cite{sornborger3}, 
symmetric configurations in the model we are 
considering are in fact unstable to nonspherical collapse. 
Nonetheless, we will continue to refer to the initial data described 
by Eq.(\ref{carrolldata}) and Eq.(\ref{ourchi}) as ``spherically 
symmetric" mainly to distinguish it from another set of initial
data which we investigate that has toroidal symmetry.  
  
\begin{figure}
\epsfxsize=7.5cm
\centerline{\epsffile{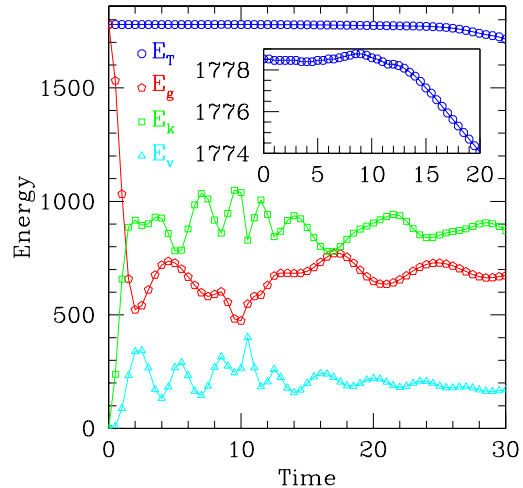}}
\caption{
The energy components for the ``spherically symmetric'' texture
as a function 
of time.
Here,  $\gamma=0.1$, $R=10$, and $\beta=3.5$ with outer boundary $z_{\rm max} = |z_{\rm min}| = \rho_{\rm max} =40$.
This run was on 513x1025 grid. The inset shows
the total energy as a function of time. The total energy is conserved
to less than one percent until gradient energy reaches the outer
boundary.
}
\label{fig:axi_carroll_es}
\end{figure}

On evolving this data, 
we find that as the texture collapses, the energies evolve as
described in~\cite{sornborger3}.  The kinetic and gradient energies evolve
towards equipartition
while the potential energy remains a relatively small fraction of the total.
Fig.~\ref{fig:axi_carroll_es} shows the energy versus time for the collapse
of a typical ``spherically symmetric" texture. 

\begin{figure}
\epsfxsize=7.5cm
\centerline{\epsffile{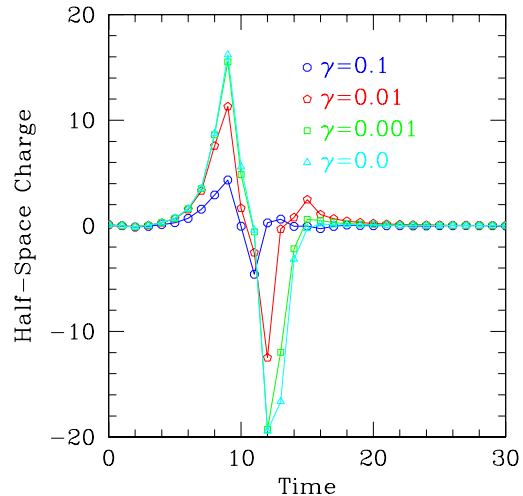}}
\caption{
The half-space charge as a function
of time. Initial data is the spherically symmetric texture
with $R=10$ and $\beta=3.0$.
Run on a 129x129x129 grid.
}
\label{fig:3d_carroll}
\end{figure}

However, our results in both the 2D axisymmetric and the 3D codes
do not bear out the claim that a monopole-antimonopole pair is
formed~\cite{sornborger3}. 
As one indication, the half-space charge of the ``spherically symmetric" 
texture as shown in Fig.~\ref{fig:3d_carroll} does
not resemble that shown for the pair in Fig.~\ref{fig:pair}.
Fig.~\ref{fig:3d_carroll}
suggests that there is a separation in the monopole charge density, but
the duration of the resulting peaks in the charge density is independent of
$\gamma$ and the constant $R$. Further, the half-space charge does not
asymptote to $\pm 1$ as one might expect were a pair to form and 
move apart.  This provides additional 
evidence that no monopoles are
forming.

\begin{figure}
\epsfxsize=7.5cm
\centerline{\epsffile{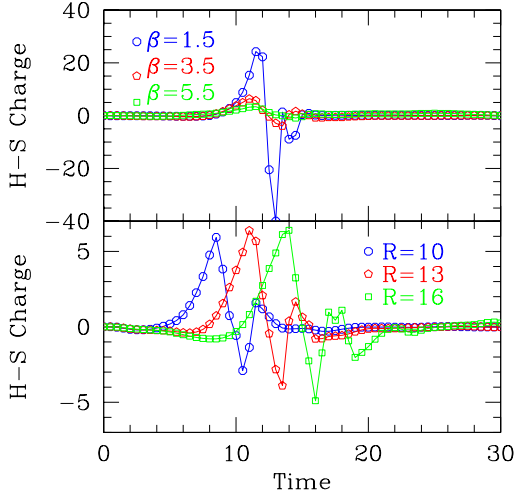}}
\caption[Half space charge for different values of
$R$ and $\beta$ for  the ``spherically symmetric'' texture.]
{Half-space charge for the ``spherically symmetric''
texture for different values of $R$ and $\beta$. The top
graph shows the results from variations of $\beta$ with $R=13$.
The bottom shows the results of variations of $R$ with $\beta=3.5$.
Here, $\gamma=0.1$ on a 513x1025 grid.
}
\label{fig:variations}
\end{figure}

The half-space charge is also plotted for variations of $R$
and $\beta$ in Fig.~\ref{fig:variations}. Again, the figure
suggests that the ``spherically symmetric" texture does not
nucleate monopole pairs.

One measure, given in~\cite{sornborger3} and~\cite{rhie}, for determining
if a pair is present is that the potential energy density should
be sharply peaked at the location of a monopole or antimonopole.
We do see some peaking in the potential (and total) energy density
for a brief period on the axis of symmetry, but the peaks do not 
persist.  Indeed, these peaks quickly diminish in size 
and propagate off the grid with the remainder of the outgoing
radiation (see further discussion in Section~\ref{sec:tex_discussion}).

Thus we confirm earlier results of Rhie and Bennett~\cite{rhie} that 
pair production does not occur with this initial data in
contradiction to that claimed in~\cite{sornborger3}.

\section{Toroidally Symmetric Texture}  
\label{sec:tor}
That the ``spherically symmetric" texture does not collapse via the
production of global monopoles does not exclude the possibility of pair production,
merely our choice of initial data.
Indeed, experiments with liquid crystals provide evidence that
monopole pairs are in fact formed in Hopf texture collapse.  
For this reason, different initial data is suggested in~\cite{luo} as
possibly leading to pair creation.  This initial data is toroidally symmetric.
It is given by 
\beq
\Phi^1 & = & \cos \theta(\eta) \\
\Phi^2 + i\Phi^3 & = & \sin \theta(\eta)~e^{i (\beta - \varphi)} 
\eeq
where, in terms of Cartesian coordinates $(x,y,z)$, the toroidal coordinates 
$(\eta, \beta, \varphi)$ are defined as  
\beq
\tanh\eta & = & {2 a \sqrt{x^2 + y^2} \over x^2 + y^2 + z^2 + a^2} \\
\tan\beta & = & {2 a z \over x^2 + y^2 + z^2 - a^2} \\
\tan\varphi & = & {y \over x}.
\eeq
The parameter $a$ is the value of the so-called degenerate tori
in this coordinate system.  
The function $\theta(\eta)$ has boundary conditions $\theta(0)=0$ 
(on the axis of symmetry {\it and} $r\rightarrow\infty$) 
and $\theta(\infty) = \pi$ (the degenerate torus).  Thus a simple
choice for $\theta$ is $\theta(\eta) = \pi \tanh(\eta)$.   
Hence, we can write the initial data as
\be
\Phi(\vec x) = \frac{1}{\sqrt{\kappa}}\left( \begin{array}{c}
                                             \cos \left( 2\pi a \frac{\sqrt{x^2+y^2}}{r^2+a^2} \right) \\
                                             \sin \left( 2\pi a \frac{\sqrt{x^2+y^2}}{r^2+a^2} \right)
                                               \frac{ \left( r^2-a^2 \right) x + 2 a z y  }
                                                    { \sqrt{ \left( \left( r^2-a^2 \right)^2 + 4a^2 z^2 \right) \left( x^2 + y^2 \right)} } \\
                                             \sin \left( 2\pi a \frac{\sqrt{x^2+y^2}}{r^2+a^2} \right)
                                               \frac{-\left( r^2-a^2 \right) y + 2 a z x  }
                                                    { \sqrt{ \left( \left( r^2-a^2 \right)^2 + 4a^2 z^2 \right) \left( x^2 + y^2 \right)} } 
                                             \end{array} \right) .
\label{eq:tor_init}
\ee

\begin{figure}
\epsfxsize=7.5cm
\centerline{\epsffile{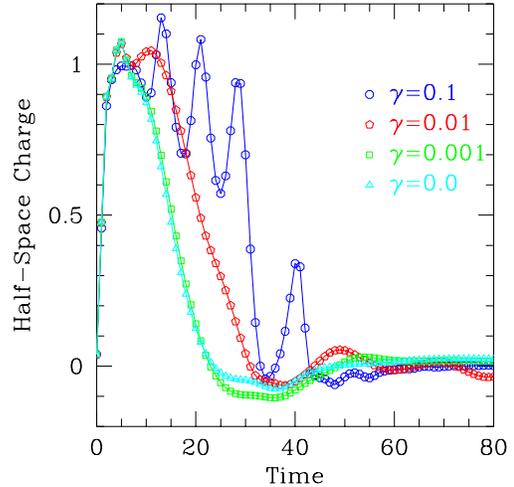}}
\caption{
The half-space charge as a function
of time. The initial data is the toroidally symmetric texture 
with $a=10.01$ and with outer boundary $z_{\rm max/min} = x_{\rm max/min} = y_{\rm max/min}= \pm 40$.
Run on a 129x129x129 grid.
}
\label{fig:3d_toroid}
\end{figure}

The evolutions of the toroidally symmetric texture generically display the
creation of a monopole-antimonopole pair for all the values of $a$ we attempted.
Fig.~\ref{fig:3d_toroid} shows the half-space charge
for a typical run. These results are strikingly similar to those
shown for the explicit pair evolved in Fig.~\ref{fig:pair}.
The charge begins at zero, quickly reaches a value approximately $\pm1$,
and remains there until collapse. As $\gamma$ is decreased,
the behavior fundamentally changes in the same manner as
that for the explicit pair. Hence, this charge separation
appears intimately connected with the potential as would
be expected for pair production. This connection is in 
contrast with the nearly $\gamma$-independent results
of the ``spherically symmetric'' texture in Fig.~\ref{fig:3d_carroll}.

Even more convincing evidence for the production of a monopole
pair resides in actual
movies
made from the evolutions.
Looking at either the potential energy density or total energy density,
two peaks representing the monopoles occur along the axis of symmetry.
The peaks leave the center of the grid, travel outward on the axis,
and eventually switch direction and annihilate.

\section{Discussion}
\label{sec:tex_discussion}

The study of topological defects is important for studies
of structure formation in cosmology,
studies of superconductivity, liquid crystals, and superfluidity
in condensed matter, and studies of regularity in
topology.

\begin{figure}
\epsfxsize=7.5cm
\centerline{\epsffile{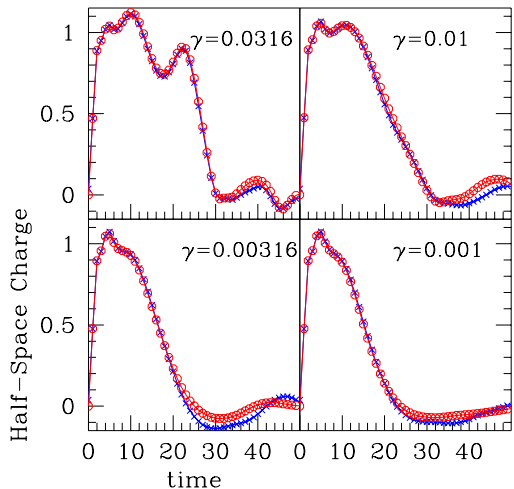}}
\caption{Comparison of the axisymmetric and three dimensional
codes using the toroidally symmetric texture as initial data.
The 2D runs (open circles) were on a 513x1025 grid while
the 3D grid (crosses) was 129x129x129.
The graph shows agreement between the two codes.
}
\label{fig:axi_3d_compare}
\end{figure}

Here, we construct a code in 3D and in axisymmetry 
to model defect dynamics. Tests of
the both codes  show that they converge with increasing
resolution and conserve energy. Further, the 3D code
duplicates the results generated with the axisymmetric code
as shown in Fig.~\ref{fig:axi_3d_compare}.

Our code clarifies some of the conditions under which a Hopf texture
nucleates a monopole-antimonopole pair. Specifically, our results
confirm those of Rhie and Bennett~\cite{rhie} and Luo~\cite{luo}
that the
``spherically symmetric'' texture does {\em not} produce monopole
pairs. This finding is in contradiction to the conclusion of~\cite{sornborger3}.
However, our results are in agreement with their conjecture 
that models of texture collapse which allow more than 
one non-trivial homotopy group will evolve differently from 
those with just non-trivial $\pi_3$.  That toroidally symmetric initial
data in this model leads to pair production agrees with their observation that 
these models can nucleate defects characterized by the non-$\pi_3$ 
homotopy group.  

We speculate that the reason monopole production was claimed
in~\cite{sornborger3}
is that the collapse of a ``spherically symmetric'' texture
does produce massive radiation in shells which are peaked along
the axis. By examining an isosurface of the potential energy,
apparent balls are seen radiating along the axis of symmetry.
However, our evolutions suggest that observing isosurfaces
can be deceiving. Indeed, by examining the charge density,
no localization is observed.

\begin{figure}
\epsfxsize=7.0cm
\centerline{\epsffile{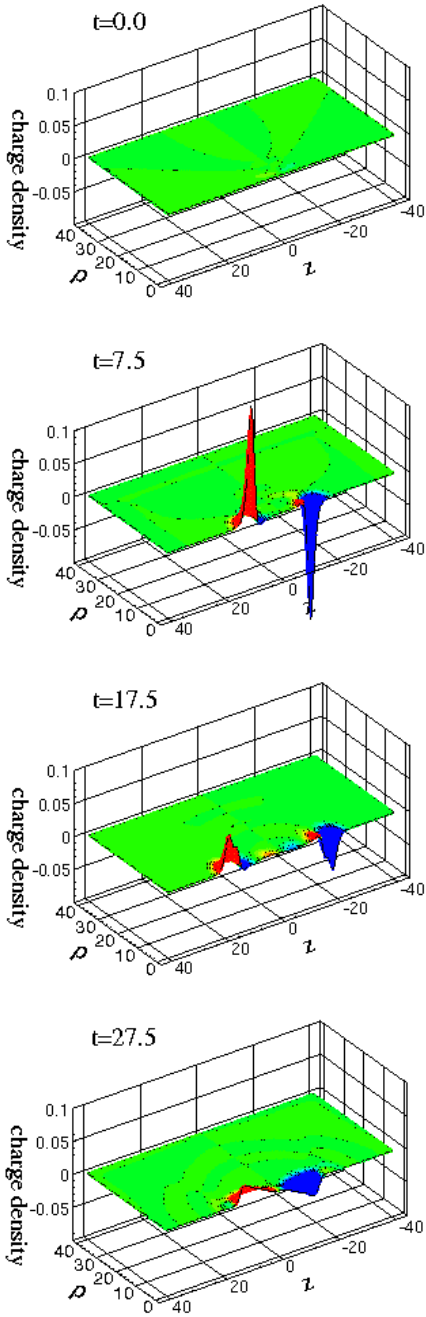}}
\caption{Evolution of the monopole charge density for
a toroidally symmetric texture ($a=10.01$). A monopole
pair is nucleated by $t=7.5$. By $t=27.5$, the pair
is accelerated towards each other about to annihilate.}
\label{fig:tecplot_tor}
\end{figure}

\begin{figure}
\epsfxsize=7.0cm
\centerline{\epsffile{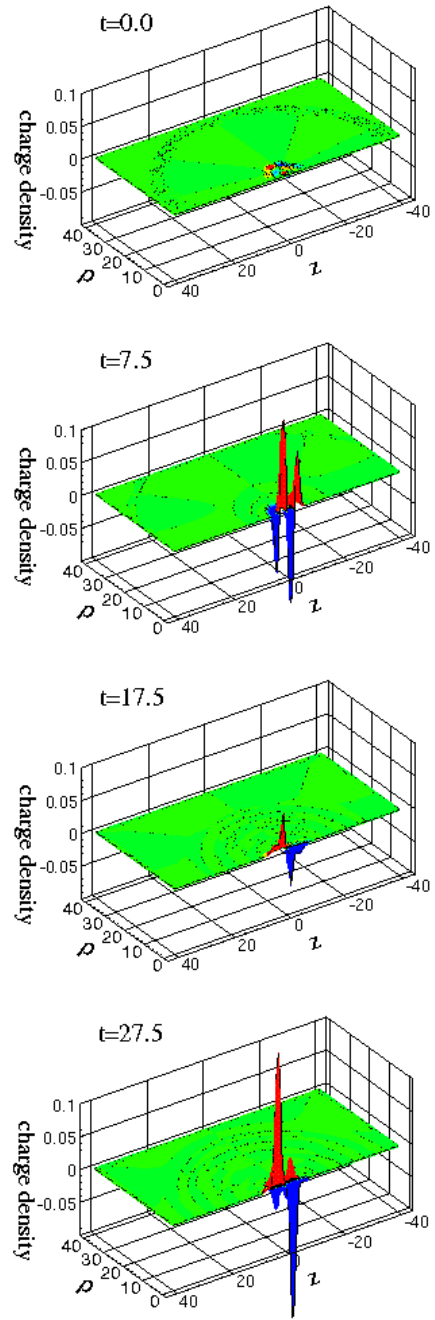}}
\caption{Evolution of the monopole charge density for
a spherically symmetric texture ($R=5$ and $\beta=3.5$).
The charge density near the origin separates such that there
are positive and negative regions. However, at no time do
these regions move apart, and instead they oscillate near the origin.
In addition, no correspondence between regions of non-zero
charge density and trapped potential energy (as shown in Fig.~\ref{fig:tecplot_ev_sph})
is observed.}
\label{fig:tecplot_sph}
\end{figure}

Consider, for example, Figs.~\ref{fig:tecplot_tor} and~\ref{fig:tecplot_sph}. 
These show the charge densities for four snapshots in time
for both the toroidally symmetric and ``spherically symmetric" 
textures evolved using the 2D axisymmetric code. 
Examining the toroidally symmetric texture first,
pair nucleation is quite dramatic between the first two
frames. From an initially vanishing charge density, two
localized, oppositely charged regions
move outward from the origin. Between the last two frames,
the monopoles change direction and move inward towards
annihilation.

Contrast these dynamics with that
shown in Fig.~\ref{fig:tecplot_sph} for the
``spherically symmetric" texture. Peaks in the charge density 
appear on the axis near the origin, but at no time are localized
regions of non-vanishing charge density seen moving
outward. Instead, the dynamics show only a region
surrounding the origin in which the charge density 
oscillates.  This is consistent with the form of the half-space charge in 
Figs.~\ref{fig:axi_carroll_es} and~\ref{fig:3d_carroll}, which shows a 
steep rise and 
fall in this quantity.  

\begin{figure}
\epsfxsize=7.0cm
\centerline{\epsffile{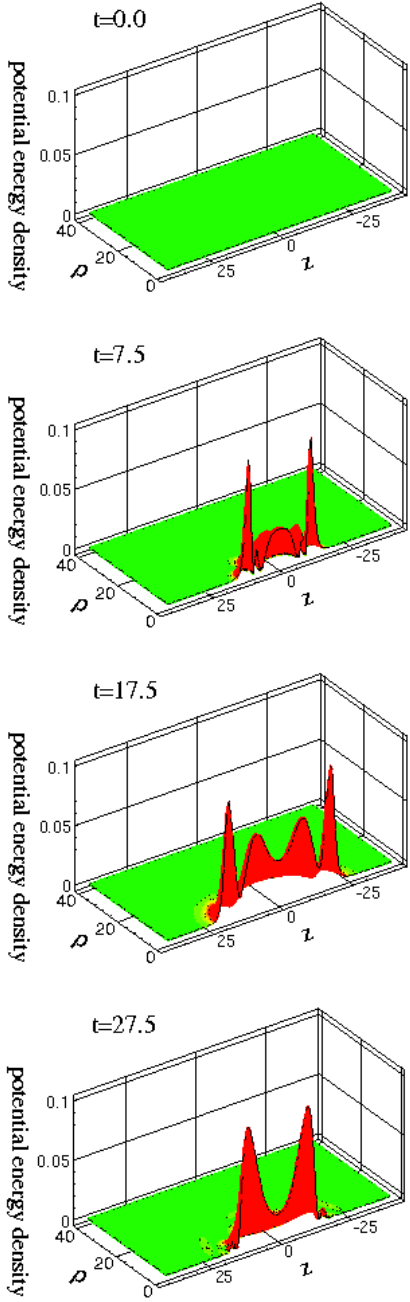}}
\caption{Evolution of the potential energy density for
the toroidally symmetric texture shown in Fig.~\ref{fig:tecplot_tor}.
At times $t=7.5$ and $t=17.5$, localized regions of trapped potential
region corresponding to the regions of localized charge density
as shown Fig.~\ref{fig:tecplot_tor} are observed. These regions
appear to represent monopoles. At $t=27.5$, the monopoles are
about to annihilate.}
\label{fig:tecplot_ev_tor}
\end{figure}

\begin{figure}
\epsfxsize=7.0cm
\centerline{\epsffile{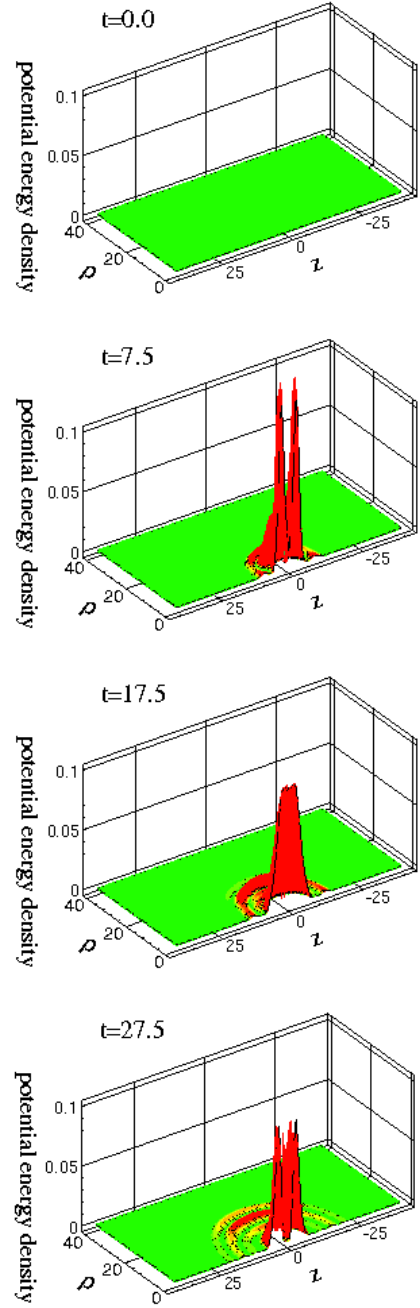}}
\caption{Evolution of the potential energy density for
the spherically symmetric texture shown in Fig.~\ref{fig:tecplot_sph}.
Potential energy is seen near the origin. Massive radiation
is seen emanating from the dynamics at the origin, however, no
trapped, localized regions which might correspond to monopoles are
observed.
}
\label{fig:tecplot_ev_sph}
\end{figure}

In addition, we show the potential energy density
associated with these evolutions in
Figs.~\ref{fig:tecplot_ev_tor} and~\ref{fig:tecplot_ev_sph}.
If monopoles exist in the space, then regions of non-zero
charge density should be accompanied by trapped potential
energy. For the toroidally symmetric texture,
Fig.~\ref{fig:tecplot_ev_tor} demonstrates
that the localized regions of non-zero charge
density in Fig.~\ref{fig:tecplot_tor}
also contain localized potential energy.  Note that these localized peaks 
in the potential energy maintain their distinct character until  
they recombine and annihilate.  Hence,
we conclude that this provides strong evidence that
these localized regions represent monopoles.

However, for the ``spherically symmetric" texture shown
in Fig.~\ref{fig:tecplot_ev_sph}, the potential energy
remains concentrated around the origin while shells of
radiation move outward.  Again, there is peaking in the potential
energy density, however these concentrations do not  
maintain any distinct form.  In particular, at times there are two
peaks and at others there is only one in the strong field region near
the origin.  Outgoing radiation can be observed (this is seen more 
clearly in the
evolutions)
 which is peaked along
the axis, but these are not accompanied by a similar peaking in the 
charge density.  
Taken together, we therefore conclude from this evidence that
no monopoles are nucleated with this initial data.

As we mentioned earlier, supplementary to these plots,  
we have produced 
movies
of some of these evolutions.  These evolutions can be viewed at
{\tt http://godel.ph.utexas.edu/\~{}ehirsch/hopf.html}.  
 
Both the axisymmetric and 3D codes are written in RNPL,
an advanced numerical language developed at the Center~\cite{marsa}.
Development of this language is ongoing, and we expect soon
to have the ability to take advantage of automatic parallelization
so that higher resolutions can be run. We also expect to adapt
these codes to model other broken symmetries.

\section*{Acknowledgments}
\label{sec:ack}
We would like to thank Matthew Choptuik
and Andrew Sornborger for helpful discussions and correspondence.
This work has been supported by NSF grants PHY9722068
and PHY9318152.  These computations were done in part
with support from a NPACI Metacenter Grant with
time on the San Diego Supercomputer Center's T90 machine.


\cleardoublepage

\end{document}